\documentstyle[12pt,dina4,epsf]{article}

\begin{document}

\title{AC--Conductivity of Pinned Charge Density Wave Fluctuations
in Quasi One--Dimensional Conductors}

\author{W. Wonneberger \\
\\
{Department of Physics, University of Ulm}\\
{D--89069 Ulm, Germany\/}}

\date{}
\maketitle

\noindent{\bf Abstract}.\quad
Quasi one--dimensional conductors which undergo a Peierls transition to a
charge
density wave state at a temperature $T_P$ show a region of
one--di\-men\-sion\-al
fluctuations above $T_P$. The Ginzburg--Landau--Langevin theory for the
frequency dependent collective conductivity from conductive fluctuations
into the charge density wave state
is developed. By inclusion of a phase breaking term the effect of local
pinning due to random impurities is simulated. It is found that the
spectral weight of the unpinned fluctuations is partly redistributed into
a pinned mode around a pinning frequency in the far infrared region.
In addition, selection rule breaking by the impurities makes the
fluctuating amplitude mode visible in the optical response.

\vspace{1cm}

\noindent
Short title: Ac--Conductivity of Pinned CDW Fluctuations

\vspace{1cm}

\\

\vspace{3cm}

\noindent
Phone ++49 731 502 2991/4\\
Fax ++49 731 502 3003\\
e--mail wonneberger@physik.uni-ulm.de\\

\newpage

\section{Introduction}

Quasi one--dimensional conductors like the transition metal chalcogenides
are characterized by a nested Fermi surface. This renders them unstable
to the Peierls transition when electron--phonon backscattering between
the two sheets of the Fermi surface become relevant. We call this a
Peierls--system (PS). The ensuing charge density wave (CDW) which develops
below a transition temperature $T_P$ of the order of $200K$ shows many
unusual properties, especially in its electronic responses \cite{G88}.

\noindent
In recent years it became increasingly evident that the electronic properties
of PS in their normal phase ($T>T_P$) are also unusual. Photoemission
studies (\cite{AGC95,GAC96} and earlier work cited therein) point towards
possible non--Fermi--liquid (NFL) behaviour. The microwave and optical
response also deviates from the Drude predictions for a normal metal \cite{
GVK94, SDA95}. Specifically, the real part of the complex conductivity
function $Re\sigma(\omega, T_0)$ for the chain direction and at room
temperture $T_0$ shows the presence of a pseudo--gap at $\omega \approx
2\Delta_0$ where $\Delta_0$ is the zero temperature half gap of the
corresponding CDW. In addition, a peak structure appears in the
far infrared well below the pseudo--gap. This peak resembles the
pinned Fr\"ohlich mode at about the same frequency but seen in the
fully developed CDW. Similar results were also found in PS films
\cite{PDL98}.

\noindent
In principle these features are known for a long time from the CDW material
$K_2P_t(CN)_4Br_{0.3}\cdot3(H_2O)$ (KCP) in which fluctuation
effects due to
intrinsic disorder are prominent. In \cite{BSZ75} the ac--conductivity
of KCP is modelled by a dielectric function which takes pinning--unpinning
fluctuations into account.

\noindent
A related explanation for the optical data is provided by the
concept of fluctuating CDW segments which behave similar to the fully
developed CDW. These segments are pinned by random impurities which break
the phase invariance of the equation for the fluctuating order parameter.

\noindent
In view of the one--dimensional nature of fluctuations above the
temperature $T^* > T_P$ below which transverse fluctuations set in to
initiate the phase transition, the observed effects could also be a signature
of NFL behaviour. The latter seems to be established for some of the
Bechgaard salts \cite{MGA98, DSG96, SDG98} where CDW fluctuations are
not important.

\noindent
In the case of PS, the attractive backscattering and the softening of the
Peierls phonons make a plain Luttinger--liquid scenario unlikely. Voit
\cite{JV95,JV96}
advocates a Luther--Emery state \cite{LE74} for the electrons in blue bronze
as explanation of the photoemission spectra. In this case a spin gap would
open before true CDW formation. Such conclusions, however, are not
undisputed. Shannon and Joyce \cite{SJ96} argue in favour of a model
\cite{LRA73,RS73} for a fluctuating Peierls system (FPS). The unusual plasmon
dispersion in quasi one--dimensional conductors can also be understood in
terms of a conventional band picture \cite{SGP98}. There seems to be good
reason to persue the FPS concept for one--dimensional metals with CDW
fluctuations.

\noindent
The FPS model has recently been extended by McKenzie \cite{McK95}. He
calculates the one--particle Green's function for Gaussian
order parameter fluctuations and large correlation length using Sadovskii's
exact method \cite{Sad}.
He points out that the electron spectral function of FPS is of NFL type
as found earlier in the same context \cite{WL76}.
McKenzie also calculates renormalized coefficients for Ginzburg--Landau
functionals of PS. The model which we will solve for the collective
conductivity corresponds to the FPS concept.

\noindent
The present paper computes the frequency dependent conductivity from
one--di\-men\-sio\-nal conductive order parameter fluctuations using a
modified, linear
Ginzburg--Landau--Langevin (GLL) equation for CDW including phase breaking
by impurities. In superconductivity (SC) where phase pinning does not exist,
this part of the conductivity was first investigated by Aslamazov and Larkin
\cite{AL68} (AL). The CDW version of the AL--theory was given in \cite{ST74}.

\noindent
From SC it is known \cite{AV93,AZ84} that there are two further contributions
from
order parameter fluctuations to the dc conductivity: A resistive contribution
from the reduction of the single particle density of states which is related
to the pseudo gap and the anomalous Maki--Thompson \cite{KM68,RT70} term
which is conductive.
In CDW the latter becomes resistive \cite{KM90} and dominates the dc
conductivity near the transition.

\noindent
The issue of one--dimensional collective dc conductivity in CDW was strongly
debated in the seventies \cite{ABB74, FRV74, ST74, PS74, TS78, Sch78}. In
\cite{ABB74} the idea of paraconductivity in one--dimensional metals with
dominant electron--phonon interaction was
advanced using a GL approach. This result was criticized in \cite{FRV74}
where pinned collective fluctuations were shown to reduce the dc
conductivity. These authors also studied the corresponding results for the
Hubbard and the Tomonaga--Luttinger model. Detailed microscopic studies of
FPS in \cite{ST74, PS74} find
both resistive and conductive fluctuations but neglect phase pinning. The
dc paraconductivity of commensurate FPS was studied in \cite{Sch78} in a GL
context. This paper served as a starting point for the present work.

\noindent
The paper is organized as follows: Sec. 2 develops the GLL approach
for CDW and derives the known results for unpinned conductive fluctuations.
Sec. 3 introduces a
modified GLL--equation where phase invariance is broken in a way which
simulates local pinning by random
impurities, and evaluates the basic correlation function of the order
parameter fluctuations. This result is used to calculate
the frequency dependent collective conductivity exactly within the model.
The complicated formula is evaluated approximately for pinning frequencies
small in comparison to the frequency of amplitude fluctuations in Sec. 4.
Two appendices present mathematical details.

\section{Ginzburg--Landau--Langevin Approach}

It is interesting to formulate the problem of fluctuation
ac--conductivity in CDW
without pinning using the GLL--method. One expects to find close
similarities to the elegant formulations for SC \cite{Ti96,RF69}. However,
it turns out that one must go beyond overdamped dynamics which
in case of CDW would only give an instantaneous response in the current
correlation function and thus a frequency independent conductivity.

\noindent
Our starting
point is the GLL equation directly in terms of the gap fluctuations
$\Delta_k$:

\begin{equation}\label{2.1}
\ddot \Delta_{k}(t)+\gamma_0 \,\dot \Delta_{k}(t) +\omega_k^2\, \Delta_{k}(t)
=\Gamma_k(t).
\end{equation}

\noindent
The parameters	$\gamma_0$ and $\omega_k$ are taken from \cite{ABB74,TS78}.
Static parameters below correspond to the rigid lattice values in \cite{McK95}.
For simplicity the renormalization of the rigid lattice (RL) values due to
fluctuations as
proposed in \cite{McK95} is not considered. Together with Hartree--Fock
corrections to the linear GL--equation as in \cite{JS82}, it would extend
the region
of applicability of the linearized approach which is marginal at the
RL--level. This and the absence of resistive fluctuations prohibit a quantitative
comparison with experiments. Below the CDW transition temperature
McKenzie's approach \cite{McK95} is neccesary to give the characteristic
optical absorption which has been measured in \cite{DGK94}.

\noindent
The damping constant $\gamma_0$ is

\begin{equation}\label{2.2}
\gamma_0 = \omega_A^2 \frac{\hbar \pi }{8 k_B T},
\end{equation}

\noindent
where

\begin{equation}\label{2.3}
\omega_A^2 = \lambda  \omega_Q^2
\end{equation}

\noindent
is the frequency of the amplitude mode of the fully developed
CDW \cite{LRA74}, $\omega_Q$ is
the bare frequency of the $2k_F$--phonon which goes soft, and $\lambda$ is
the electron--phonon coupling constant.

\noindent
The actual frequency $\omega_k$ of the amplitude fluctuations in (\ref{2.1})
is

\begin{equation}\label{2.4}
\omega_k^2 = \omega_0^2 \left( 1 + k^2 \xi ^2 \right).
\end{equation}

\noindent
with

\begin{equation}\label{2.5}
\omega_0^2 = \omega_A^2 \epsilon_{RL}.
\end{equation}

\noindent
For definiteness we assume underdamping ($\gamma_0<2\omega_0$) which is
the case in the linear fluctuation regime sufficiently above the
transition temperature.

\noindent
Without a form of selection rule breaking neither $\omega_A$ nor $\omega_0$
can be observed in optical conductivity measurements. The amplitude mode
$\omega_A$ can be seen, however, in Raman scattering \cite{SLH75, TMW83}. The
fluctuating amplitude mode above the transition temperature is observed
in neutron scattering studies \cite{CSW76, PHS91}.

\noindent
The correlation length $\xi$ in (\ref{2.4}) is

\begin{equation}\label{2.6}
\xi ^2 = \xi _0^2 / \epsilon_{RL},
\end{equation}

\noindent
with $\epsilon_{RL}$ given by

\begin{equation}\label{2.7}
\epsilon_{RL}  = \ln \frac{T}{T_{RL}},
\end{equation}

\noindent
where $T_{RL}$ is the rigid lattice mean field transition temperature. The
re\-fer\-ence length $\xi_0$ is given by \cite{TS78}

\begin{equation}\label{2.8}
\xi _0^2 = \frac{7\zeta (3) \hbar^2 v_F^2}{16\pi ^2 (k_BT)^2}.
\end{equation}

\noindent
The complex Gaussian Langevin force $\Gamma$ has zero mean. Its correlation
function

\begin{equation}\label{2.9}
\langle\Gamma_k(t)\Gamma^*_{k'}(0)\rangle=
\langle |\Delta _k|^2 \rangle_0 \,2 \gamma_0 \omega_k^2\delta_{k,k'}\,\delta(t)
= 2 \gamma_0 \omega_0^2\,\frac{k_B T}
{f_0} \,\delta_{k,k'}\,\delta(t)\equiv A\,\delta_{k,k'}\,\delta(t).
\end{equation}

\noindent
is constructed in such a way as to give the fluctuation intensity

\begin{equation}\label{2.10}
\langle |\Delta _k|^2 \rangle_0 = \frac{k_B T}{a_k}.
\end{equation}

\noindent
The latter follows from the linear free energy functional

\begin{equation}\label{2.11}
F_0 = \sum _k a_k |\Delta_k|^2,
\end{equation}

\noindent
with

\begin{equation}\label{2.12}
a_k = f_0 \left( 1 + k^2 \xi ^2 \right), \quad
f_0 = \frac{ L \epsilon_{RL} }{\pi  \hbar v_F}.
\end{equation}

\noindent
Note that (\ref{2.1}) implies a spatial correlation of the order
parameter according to

\begin{equation}\label{2.13}
\langle \Delta(x,0)\Delta^*(0,0)\rangle =\frac{L}{2\pi}\int\,dk\, e^{ikx}
\langle|\Delta_k|^2\rangle_0=\left(\frac{k_BT\pi\hbar v_F}{2\xi
\epsilon_{RL}}\right)\,e^{-|x|/\xi}\equiv \psi_{RL}^2 \,e^{-|x|/\xi}.
\end{equation}

\noindent
In the next step the one--dimensional conductivity is computed from
the classical Kubo formula

\begin{equation}\label{2.14}
\sigma  (\omega ) = \frac{L}{k_BT}\, \int^\infty _0 \, dt e^{i\omega
t}\langle J(t)J(0) \rangle,
\end{equation}

\noindent
where $L$ is the sample length.

\noindent
The collective current density was calculated in \cite{Sch78}
\footnote{S.N. Artemenko informed the author that he obtained the same
expression for the collective current density by the Keldysh
approach to CDW dynamics.} and reads:

\begin{equation}\label{2.15}
j (x,t) = i \,\frac{b}{2} \,(\dot{\Delta }(x,t) \Delta^*(x,t)
- \Delta(x,t) \dot{\Delta}^*(x,t)).
\end{equation}

\noindent
The collective current is proportional to the time derivative $\dot{\varphi}$
of the order
parameter phase as for the fully developed CDW but the prefactor is different.
The coefficient $b$ in (\ref{2.15}) is \cite{TS78}

\begin{equation}\label{2.16}
b^2 = \left( \frac{e_0}{2k_BT\hbar \nu _b} \right)^2,
\end{equation}

\noindent
involving the backward scattering rate $\nu_b$ due to random static scattering
centers. This formula holds in the pure limit when the electron scattering
rate obeys $\hbar \nu \equiv \hbar(\nu_f+\nu_b/2) < 2 \pi k_B T$.

\noindent
The homogeneous current density $J$ in (\ref{2.15}) is related to $j$ by

\begin{equation}\label{2.17}
J(t) = \frac{1}{L} \int^L_0 \, dx j (x,t) = j_{k=0}(t).
\end{equation}

\noindent
In the linear setting of (\ref{2.1}) not only $\Gamma$ obeys Gaussian
statistics but also $\Delta$ and exact Gaussian decoupling gives

\begin{eqnarray}\label{2.18}
\langle J (t) J (0) \rangle = \frac{b^2}{2} \sum_k
\left[ \dot{C}(k,t)^2 - C(k,t) \ddot{C}(k,t) \right],
\end{eqnarray}

\noindent
provided the correlation function

\begin{equation}\label{2.19}
C(k,t) \equiv \langle \Delta_k (t) \Delta _k^ * (0) \rangle
\end{equation}

\noindent
is real and even in t. $C(k,t)$ is evaluated from (\ref{2.1}) and
explicitly given by

\begin{equation}\label{2.20}
C \left( k, t \right) = \langle |\Delta _k |^2 \rangle_0 \exp(
 - \frac{\gamma _0}{2} |t| )
\left[\cos D_k t+ \frac{\gamma_ 0}{2 D_k} \sin D_k |t| \right],
\end{equation}

\noindent
with

\begin{equation}\label{2.21}
D_k =  \sqrt{\omega_k^2 - \frac{\gamma _0^2}{4}}.
\end{equation}

\noindent
This leads to

\begin{eqnarray}\label{2.22}
\dot{C}^2 - C \ddot{C} = \langle |\Delta _k|^2 \rangle^2_0\,\,
\omega_k^2 \, \exp(-\gamma_0 |t|).
\end{eqnarray}

\noindent
Note that all oscillating terms in the correlation functions
cancel out leaving a purely relaxational response.
If one were to use the correlation function

\begin{equation}\label{2.23}
C(k,t) = \langle |\Delta _k|^2 \rangle_0 \, \exp(-\gamma_k |t|),
\end{equation}

\noindent
for the overdamped version of (\ref{2.1})
with $\gamma_k=\omega_k^2/\gamma_0$ one would get an instantaneous
response $ \langle J (t) J (0) \rangle \propto \delta(t) $ and hence a
frequency independent conductivity.

\noindent
Calculation of the conductivity using (\ref{2.22}) gives, however, the
correct result given in \cite{JS82}

\begin{equation}\label{2.24}
Re\, \sigma(\omega) = \,\sigma _F\, \frac{\gamma_0
^2}{\gamma _0^2 + \omega ^2},
\end{equation}

\noindent
irrespective of the relation between $\gamma_0$ and $\omega_0$.
The scale--value $\sigma_F$ of the fluctuation conductivity
is

\begin{equation}\label{2.25}
\sigma_F= \frac{L^2 A^4 b^2}{16 k_BT \omega_0^2 \xi \gamma_0^3},
\end{equation}

\noindent
and coincides with the result  \cite{TS78}. Explicitly $\sigma_F$ reads

\begin{equation}\label{2.26}
\sigma_F=\frac{2\pi^2e_0^2k_BT v_F}{\sqrt{7 \zeta(3) \epsilon_{RL}}
\,\,(\hbar\nu_b)^2}.
\end{equation}

\noindent
The conductivity shows the
mean--field critical behaviour $\sigma_F \propto \epsilon_{RL}^{-1/2}$.
In the picture of a metal with order parameter fluctuations
the collective conductivity adds to the normal state conductivity
$\sigma_N = 8 e_0^2 v_F/(4\pi \hbar \nu_b)$.

\noindent
Formal calculations in higher spatial dimensions require a momentum
cut--off in contrast to SC. This is related to the form of the
collective current density (\ref{2.15}).

\section{Breaking of Phase Invariance}

The space--time version of (\ref{2.1}) is

\begin{equation}\label{3.1}
\ddot \Delta(x,t)+\gamma_0 \,\dot \Delta(x,t) +\omega_0^2\,
\left(1- \xi^2\,\frac{\partial^2}{\partial x^2}\right)\,\Delta(x,t)
=\Gamma(x,t).
\end{equation}

\noindent
The simplest way to break the phase invariance of this equation is to
add a pinning term

\begin{equation}\label{3.2}
2\,\omega_i^2 \,|\Delta(x,t)|\, \cos\varphi(x,t),
\end{equation}

\noindent
to the left hand side which is a simple local coupling.
This is clearly not the general starting point to treat pinning by
random impurities \cite{FL78, LR79}. However, it will
become evident later that this approach simulates local pinning because
the final pinning frequency is proportional to the impurity concentration.

\noindent
To arrive at (\ref{3.2}) we start from the more general form

\begin{equation}\label{3.3}
\omega_s^2 \,\sum_i\,h(x-x_i)\,(\Delta(x_i,t)+\Delta^*(x_i,t)).
\end{equation}

\noindent
The impurities are locally coupled to the order parameter.
The real structure function $h(x)$ transmits the effect to
the order parameter at $x$. The terms $\Delta^*(x_i,t)$ break phase
invariance by modelling backward scattering. The scale frequencies $\omega_s$
and $\omega_i$ are different from the final pinning frequency.

\noindent
We make two further asumptions: The function $h$ is a contact interaction
$h(x)=l_i \delta(x)$ with a scattering length $l_i$. The crudest assumption
is, however,

\begin{equation}\label{3.4}
\sum_i \rightarrow n_i\int\,dx_i,
\end{equation}

\noindent
where $n_i$ is the density of impurities. This requires $ n_i \xi > 1$ and
amounts to an early impurity average. Introducing the scale frequency

\begin{equation}\label{3.5}
\omega_i^2 \equiv \omega_s^2\,n_i\,l_i
\end{equation}

\noindent
then leads to (\ref{3.2}).

\noindent
This admittedly crude model has the advantage to allow for an exact solution.

\noindent
The complete GLL--equation which replaces (\ref{2.1}) can be written:

\begin{equation}\label{3.6}
\ddot \Delta_{k}(t)+\gamma_0 \,\dot \Delta_{k}(t) +\omega_k^2\, \Delta_{k}(t)
+\omega_i^2 \, (\Delta_k(t)+\Delta^*_{-k}(t)) =\Gamma_k(t).
\end{equation}

\noindent
The complex order parameter $\Delta(x,t)$ is decomposed into real and imaginary
parts $U(x,t)= Re \Delta(x,t)$ and $V(x,t)=Im\Delta(x,t)$ giving

\begin{equation}\label{3.7}
\Delta _k = U_k + i V_k,
\end{equation}

\noindent
with complex $U_{k}$ and $V_k$ which satisfy the usual reality conditions

\begin{equation}\label{3.8}
U_{-k} = U_k^* ,\quad V_{-k} = V_k^*.
\end{equation}

\noindent
The Langevin equation splits into two equations
for $U_k$ and $V_k$:

\begin{eqnarray}\label{3.9}
\ddot U_{k}(t)+\gamma_0 \,\dot U_{k}(t) +\omega_k^2\, U_{k}(t)
+2 \omega_i^2 \, U_{k}(t) &=&\Gamma_{Uk}(t),
\\[4mm]\nonumber
\ddot V_{k}(t)+\gamma_0 \,\dot V_{k}(t) +\omega_k^2\, V_{k}(t)
&=& \Gamma_{Vk}(t).
\end{eqnarray}

\noindent
Here $\Gamma_{Uk}(t)$ and $\Gamma_{Vk}(t)$ are the Fourier transforms of
the real and imaginary part of $\Gamma(x,t)$, respectively.
It is possible that the impurities modify
the thermal random force $\Gamma(x,t)$. In our model we assume that this
is not the case. This assumption is reasonable for $\omega_i \ll \omega_0$.
The random forces $\Gamma_{Uk}$ and $\Gamma_{Vk}$ are then independent
Langevin forces with the same statistical properties as $\Gamma_k(t)$
(cf. (\ref{2.9})) but only half its strength.
The two equations  (\ref{3.9}) become independent and are both isomorphic
with (\ref{2.1}). However, the frequencies for the $U$--modes are modified
and change their fluctuation intensities:

\begin{eqnarray}\label{3.10}
\langle |U_k|^2\rangle=(1+2\frac{\omega_i^2}{\omega ^2_k})^{-1}
\frac{k_BT}{2a_k},\quad
\langle |V_k|^2\rangle=\frac{k_BT}{2a_k}.
\end{eqnarray}

\noindent
Hence the intensity of the fluctuating order parameter is reduced:

\begin{eqnarray}\label{3.10a}
\langle |\Delta_k|^2\rangle = \langle |\Delta_k|^2\rangle_0
\frac{1+ (\omega_i/\omega_k)^2}{1+ 2(\omega_i/\omega_k)^2}.
\end{eqnarray}

\noindent
A thermodynamic derivation for (\ref{3.10}) is given in Appendix A. In view
of the realistic condition $\omega_i \ll \omega_k$ the renormalization of
the mean square order parameter is irrelevant.

\noindent
The order parameter correlation function  becomes

\begin{eqnarray}\label{3.11}
C(k,t) = p _+ C_+(k,t) + p _- C_-(k,t),
\end{eqnarray}

\noindent
with weights

\begin{eqnarray}\label{3.12}
p_+ = \frac{1}{2 (1 + 2(\omega ^2_i/\omega^2_k))}, \quad p_-=\frac{1}{2}.
\end{eqnarray}

\noindent
and the replacement

\begin{eqnarray}\label{3.13}
D_k \rightarrow D_k^{(+)} = \sqrt{\omega^2_k +2 \omega_i^2-\frac{\gamma^2_0}
{4}} \equiv \sqrt{(\omega_k^{(+)})^2-\frac{\gamma^2_0}{4}}
\end{eqnarray}

\noindent
in the expression (\ref{2.20}) for $C(k,t)$ in order to get $C_+(k,t)$
while $C_-(k,t)$ remains unchanged, i.e. formally $D_k^{(-)}=D_k$,
$\omega_k^{(-)}=\omega_k$.
This solves completely the GLL--equation (\ref{3.6}).

\section{Discussion of Fluctuation Conductivity}

We define the wave--number dependent pinning frequency

\begin{equation}\label{4.1}
\omega_p(k) \equiv \frac{\omega_i^2}{D_k}.
\end{equation}

\noindent
From (\ref{3.5}) it is seen that $\omega_p(k)$ is proportional to the
linear impurity concentration. Thus our model simulates local pinning.

\noindent
Using the condition $\omega_i \ll D_k$ the result of Appendix B leads
to the following expression for the real part of the fluctuating
conductivity

\begin{eqnarray}\label{4.2}
Re  \sigma (\omega ) = \frac{L}{k_B T} \frac{b^2}{4} \sum_k
\langle |\Delta _k|^2 \rangle_0^2 \omega_k^2
\left\{
\left[\frac{\gamma_0}{\gamma_ 0^2 + \omega^2}\right]_{AL}
\right. \\[4mm]\nonumber
\left.+ \left[\gamma_0 \frac{\gamma_0^2 + \omega_p^2 (k)+\omega^2}
{\left( \gamma_0^2
+\omega^2+\omega_p^2 (k)\right)^2 -4\omega ^2 \omega_p^2(k)}
\right.\right]_P \\[4mm] \nonumber
\left.
+ \left[ \gamma_0 \frac{\omega_p^2 (k)}{4 D_k^2}
\frac{(3 - \gamma_0^2/\omega _k^2)(\gamma_0^2 + 4D_k^2)-\omega^2}
{\left( \gamma_0^2 + \omega ^2 + 4 D_k^2 \right)^2 - 16 \omega
^2 D_k^2}
\right]_A
\right\}.
\end{eqnarray}

\noindent
Even before the $k$--summation is performed three different contributions
to the fluctuation conductivity can be discriminated: the relic of the
AL--conductivity (AL) centered at zero frequency, a pinned mode (P) near the
frequency $\omega_p(0)$, and a weak structure (A) associated with the
fluctuating amplitude
modes $\omega_k$. The latter results from selection rule breaking by the
impurities. Thus traces of the fluctuating amplitude mode should be seen
in the optical
conductivity. A similar case regarding the pinned Fr"hlich mode is found
in the fully developed CDW \cite{L87, WW96}.

\noindent
The $k$--summation is easily done for the AL--part and gives

\begin{equation}\label{4.3}
Re\, \sigma(\omega)_{AL} = \frac{1}{2} \,\sigma _F\, \frac{\gamma_0
^2}{\gamma _0^2 + \omega ^2}.
\end{equation}

\noindent
This is exactly half the result (\ref{2.24}). The spectral weight $W$
according to

\begin{equation}\label{4.4}
W \equiv \int_{-\infty}^\infty d\omega	Re\, \sigma (\omega)
\end{equation}

\noindent
is

\begin{equation}\label{4.5}
W_{AL} = \frac{\pi}{2}\, \gamma _0 \sigma _F.
\end{equation}

\noindent
A lenghty but exact calculation gives for the $P$--mode

\begin{equation}\label{4.6}
Re\, \sigma(\omega)_P = \sigma _F\,[J_1(\omega) + 2 Re\,J_2(\omega)],
\end{equation}

\noindent
with

\begin{equation}\label{4.7}
J_1(\omega) = \frac{1}{2}\gamma^2_0\, \frac{\gamma ^2_0+\omega^2 -4\omega
_i^4/\gamma _0^2}{\left( \omega ^2+ \left[ \gamma _0+
2 \omega_i^2/\gamma _0
\right] ^2 \right) \left( \omega ^2+ \left[\gamma
_0-2 \omega_i^2/\gamma _0 \right] ^2 \right)},
\end{equation}

\noindent
and

\begin{equation}\label{4.8}
J_2(\omega) = 2\, \frac{\omega _0}{\gamma _0}\, \frac{\omega_i^4}{4 \omega
_i^4/\gamma _0^2- (\gamma _0+i\omega )^2}\,\,
\frac{1}{\sqrt{\left(4\omega _0^2 - \gamma _0^2 \right) \left(\gamma
_0 + i \omega  \right)^2 + 4 \omega_i^4}}.
\end{equation}

\noindent
Note the non--algebraic structure of the conductivity due to $J_2$.
The spectral weight associated with $Re\,\sigma(\omega)_P$ is independent of
pinning parameters and given by

\begin{equation}\label{4.9}
W_P = \frac{\pi}{2}\, \gamma _0 \sigma _F.
\end{equation}

\noindent
It adds the
missing half to the total spectral weight $ \pi \gamma_0
\sigma_F$ of the unpinned fluctuation conductivity.

\noindent
The amplitude mode is treated approximately. Assuming $\gamma_0 \ll
2 \omega_0$ and retaining the $k$--dependence only in the prefactor
it is found

\begin{equation}\label{4.10}
Re\, \sigma(\omega)_A =  \frac{3}{64} \frac{\omega_i^4}{\omega _0^4}\, \sigma
_F
\left\{ \frac{\gamma _0^2 \left( 12 \omega^2_0 - \omega^2
\right)}{\left( \omega ^2 - 4 \omega_0^2 \right)^2 + 4 \omega
^2\gamma_0
^2} \right\}.
\end{equation}

\noindent
The amplitude mode has a peak near $2\omega_0$. Its spectral weight
$W_A$ is small and given by

\begin{equation}\label{4.11}
W_A = \left(\frac{3}{64} \frac{\omega_i^4}{\omega _0^4}\right)\, \pi
\gamma_0 \sigma _F.
\end{equation}

\noindent
Fig. 1 shows the pinned fluctuation conductivity in comparison with the
unpinned case neglecting the weak amplitude mode.

\section{Summary}

The Ginzburg--Landau--Langevin method is developed for the fluctuation
conductivity in charge density wave systems above the transition
temperature when fluctuations are one--dimensional. An additional
phase breaking term due to impurities is introduced and its consequences
for the fluctuation conductvity is evaluated. It is found that the
spectral weight of the unpinned fluctuations is partly redistributed into
a pinned mode around a pinning frequency in the far infrared.
as seen in experiments. In addition, selection rule breaking by the
impurities enables traces of the fluctuating amplitude mode to appear
in the optical response.

\section*{Acknowledgement}
The author thanks H. Monien for helpful discussions.

\appendix

\section{Appendix: Pinned Fluctuation Intensity}

\newcounter{affix}
\setcounter{equation}{0}
\setcounter{affix}{1}
\renewcommand{\theequation}{\alph{affix}\arabic{equation}}

\noindent
The deterministic part of the GLL--equation (\ref{3.6}) can be expressed as
in terms of real variables $x_{k\nu}$ as

\begin{equation}\label{A1}
\ddot{x}_{k\nu} + \gamma _0 \dot{x}_{k\nu} = -	\frac{\omega
_k^2}{2a_k} \,\frac{\partial F}{\partial x_{k\nu}},
\end{equation}

\noindent
with the energy functional

\begin{equation}\label{A2}
F = \sum_{k,\nu} a_{k\nu} x_{k\nu}^2,
\end{equation}

\noindent
with

\begin{eqnarray}\label{A3}
a_{k\nu}=\Big{\{}
\begin{array}{ll}
a_k(1+2 b(k)),\quad \nu=1,2,\\
a_k,\quad \nu=3,4,\\
\end{array}
\end{eqnarray}

\noindent
and

\begin{equation}\label{A4}
b(k) \equiv \frac{\omega_i^2}{\omega _k^2}.
\end{equation}

\noindent
We have used the decomposition (\ref{3.7}) and splitted the components $U_k$
and $V_k$ into real and imaginary parts:

\begin{equation}\label{A5}
U_k = x_{k1} + i x_{k2} , \quad V_k = x_{k3} + i x_{k4}.
\end{equation}

\noindent
Though in our model the random forces generate no dependence among the
$x_{k\nu}$, the latter are not independent since the
$x_{k\nu}$ are even under $k \rightarrow -k$ for $\nu =1, 3$ and odd
for $\nu =2, 4$. In terms of independent $x_{k\nu}$ the
energy $F$ becomes

\begin{equation}\label{A6}
F =  2 \sum_{k>0} \,  \sum_{\nu =1}^4 a_{k \nu} x_{k\nu}^2
\equiv \sum_{k>0} \,  \sum_{\nu =1}^4 F_{k\nu}.
\end{equation}

\noindent
The statistical average of the independent variables is simply

\begin{equation}\label{A7}
\langle x_{k\nu}^2\rangle = \frac{\int dx_{k\nu} x_{k\nu}^2
\exp(-\beta F_{k\nu})}{\int dx_{k\nu}\exp(-\beta F_{k\nu})},
\end{equation}

\noindent
and gives the results (\ref{3.10}).

\section{Appendix: Calculation of Fluctuation Conductivity}

\setcounter{equation}{0}
\setcounter{affix}{2}
\renewcommand{\theequation}{\alph{affix}\arabic{equation}}

\noindent
We use (\ref{3.11}) in the Kubo formula

\begin{eqnarray}\label{B.1}
\sigma (\omega) = \frac{L}{k_B T} \frac{b^2}{2}\,\int^\infty _0 dt \,
e^{i\omega t}
\sum_k \left[ \dot{C}(k,t)^2 - C(k,t) \ddot{C}(k,t) \right].
\end{eqnarray}

\noindent
Splitting the correlation function $C(k,t)$
into its constituents gives

\begin{eqnarray}\label{B.2}
\dot{C}^2 - C \ddot{C}  = \sum_{\nu =\pm} p^2_\nu   \left(
\dot{C}^2_\nu   - C_\nu
\ddot{C}_\nu  \right) +p_+ p_- \left\{ 2\dot{C}_+ \dot{C}_-  -
C_+ \ddot{C}_-	-C_- \ddot{C}_+ \right\}.
\end{eqnarray}

\noindent
The result (\ref{2.22}) translates into

\begin{eqnarray}\label{B.3}
\dot{C}^2_\nu - C_\nu \ddot{C}_\nu  = \langle |\Delta _k|^2 \rangle_0^2
(\omega_k^{(\nu)})^2 \, \exp(-\gamma_0 |t|),\quad \nu=\pm.
\end{eqnarray}

\noindent
Hence

\begin{eqnarray}\label{B.4}
\dot{C}^2 - C \ddot{C} = \langle |\Delta _k |^2 \rangle_0^2
\exp \left(-\gamma _0 |t| \right)p_+
\left[ \omega^2_k +\omega_i^2 +\frac{1}{2}
\left \{ 2 \dot{C}_+ \dot{C}_- - C_+ \ddot{C}_- - C_- \ddot{C}_+
\right\} \right].
\end{eqnarray}

\noindent
A somewhat tedious calculation gives
a formally exact expression of the frequency dependent fluctuation
conductivity:

\begin{eqnarray}\label{B.5}
\sigma (\omega) = \frac{L}{k_B T} \frac{b^2}{2} \int^\infty_0 dt
\exp(i\omega t - \gamma_0 t)\, \sum_k
p_+\langle|\Delta _k|^2\rangle_0^2
\\[4mm]\nonumber
\left[
\omega_k^2 + \omega_i^2+\frac{1}{2}
\left(
\left\{ \frac{\omega_k^4 + 2\omega_k^2\omega_i^2 -
\gamma_0^2(\omega_k^2+\omega_i^2)/4}
{D_k^{(+)}D_k^{(-)}} \right\}
\right.\right.\\[4mm]\nonumber
\left.\left.
\left( \cos ( D_k^{(+)} - D_k ^{(-)}) t - \cos ( D_k^{(+)} +
D_k^{(-)})t\right)
\right.\right.\\[4mm]\nonumber
\left.\left.
+ (\omega_k^2+\omega_i^2) \left(\cos ( D_k ^{(+)} - D_k ^{(-)} )t +
\cos( D_k^{(+)} + D_k^{(-)}) t\right)
\right.\right.\\[4mm]\nonumber
\left.\left.
- \frac{\gamma_0 \omega_i^2}{2} \left( \frac{1}{D_k^{(+)}}
+\frac{1}{D_k^{(-)}} \right) \sin ( D_k^{(+)} - D_k^{(-)} )|t|
\right.\right.\\[4mm]\nonumber
\left.\left.
- \frac{\gamma_0 \omega_i^2}{2} \left( \frac{1}{D_k^{(+)}} -
\frac{1}{D_k^{(-)}} \right) \sin ( D_k^{(+)} + D_k^{(-)}) |t|\right)_B
\right].
\end{eqnarray}

\noindent
This result is too complicated to be discussed in full generality.
We will take advantage of the fact that in practice the condition
$\omega_i^2 \ll D_k^2$ is fulfilled and perform an expansion of (\ref{B.5})
with respect to

\begin{equation}\label{B.6}
\frac{\omega_i^2}{D_k^2} \ll 1.
\end{equation}

\noindent
This gives

\begin{eqnarray}\label{B.7}
{\left( \right)_B} &\rightarrow& 2 (\omega_k^2+\omega_i^2)
\cos \frac{\omega_i^2}{D_k}t
+ \omega_i^4 \,\frac{\omega_k^2 - \gamma_ 0^2/2}{2 D_k^4}
\cos 2D_k t
 \\[4mm]\nonumber
&-& \frac{\gamma_0 \omega_i^2}{D_k} \sin \frac{\omega_i^2}{D_k} |t|
+ \frac{\gamma_0 \omega_i^4}{2D_k^3} \sin 2D_k |t|
- \omega_i^4 \, \frac{\omega_k^2 - \gamma_ 0^2/2}{2D_k^4} \cos
\frac{\omega_i^2}{D_k} t.
\end{eqnarray}

\noindent
It is easier to do the time integral for the real part of the conductivity
in the approximated version of (\ref{B.5}).
The imaginary part follows from the Kramers--Kronig
relation. The relevant terms up to order $\omega_p^2(k)$ but neglecting
small corrections of numerical constants of order $\omega_i^2$ are
given as (\ref{4.2}) in section 4.

\begin{figure}[ht]
\epsfxsize=240pt
\epsffile{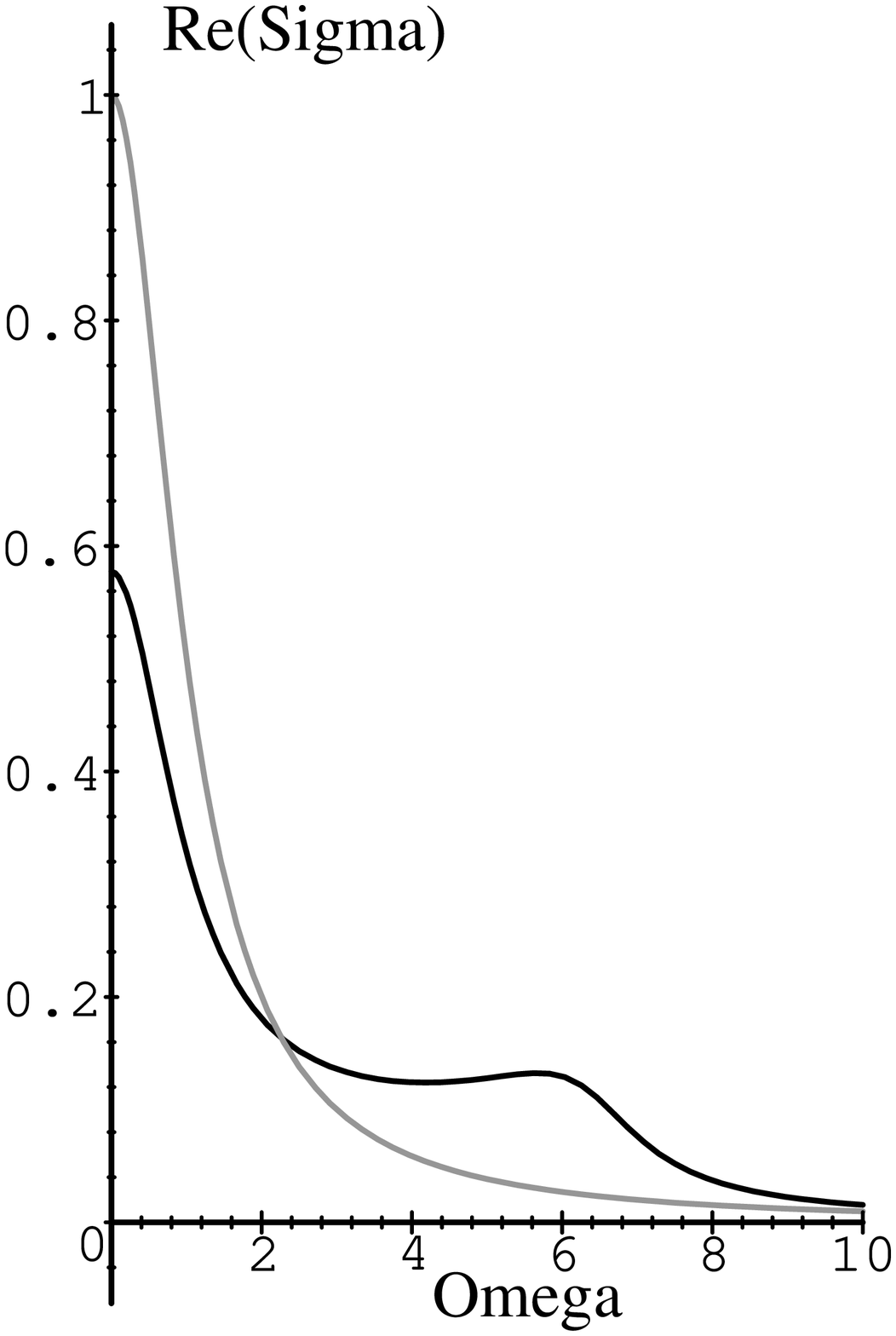}
\caption{Real part of scaled fluctuation conductivity (sum of
	 eq. (43) and eq. (46)) in comparison to
	 the unpinned case (gray line: eq. (24)) as function
	 of scaled frequency. Conductivity unit is $\sigma_F$
	 according to eq. (25) and frequency unit is the damping
	 constant $\gamma_0$ (cf. eq. (2)) of the fluctuating
	 amplitude mode. In this unit the following values were
	 chosen: amplitude mode frequency $\omega_0=40$ and pinning
	 scale frequency $\omega_i=16$. The actual pinning frequency
	 is seen to be near $6\gamma_0$.}
\end{figure}


\newpage

\begin{thebibliography}{11111}

\bibitem{G88} Gr{\"u}ner G 1988 {\it Rev. Mod. Phys.} {\bf 60} 1129

\bibitem{AGC95} Allen J W, Gweon G--H, Claessen R and Matho K 1995
{\it J. Phys. Chem. Solids} {\bf 56} 1849

\bibitem{GAC96} Gweon G--H, Allen J W, Claessen R, Clack J A,
Poirier D M, Benning P J,
Olson C G, Ellis W P, Zhang Y--X, Schneemeyer L F, Markus J,
and Schlenker C 1996 {\it J. Phys: Condensed Matter} {\bf 8} 9923

\bibitem{GVK94} Gorshunov B P, Volkov A A, Kozlov G V, Degiorgi
L, Blank A, Csiba F, Dressel M, Kim Y, Schwartz A and Gr\"uner G
1994 {\it Phys. Rev. Lett.} {\bf B 73} 308

\bibitem{SDA95} Schwartz A, Dressel M, Alavi B, Blank A, Dubois
S, Gr\"uner G, Gorshunov B P, Volkov A A, Kozlov G V, Thieme S, Degiorgi
L and Levy F 1995 {\it Phys. Rev.} {\bf B 52} 5643

\bibitem{PDL98} Pronin A V, Dressel M, Loidl A, van der Zant
H S J, Mantel O C and Dekker C 1998 {\it Physica} {\bf B244} 103

\bibitem{BSZ75} Bruesch P, Str"ssler S and Zeller H R 1975 {\it Phys. Rev.}
{\bf 12} 219

\bibitem{MGA98} Moser J, Gabay M, Auban--Senzier P, Jerome D,
Bechgaard K and Fabre J M 1998 {\it Eur. Phys. J.} {\bf B 1} 39

\bibitem{DSG96} Dressel M, Schwartz A, Gr\"uner G and Degiorgi L 1996
{\it Phys. Rev. Lett.} {\bf 77} 398

\bibitem{SDG98} Schwartz A, Dressel M, Grner G, Vescoli V, Degiorgi L
and Giamarchi T 1998 {\it Phys. Rev.} {\bf 58} 1261

\bibitem{JV95} Voit J 1995 {\it Rep. Prog. Phys.} {\bf 58} 977

\bibitem{JV96} Voit J 1996 {\it J. Phys.: Condensed Matter} {\bf 8} L779

\bibitem{LE74} Luther A and Emery V J 1974 {\it Phys. Rev. Lett.} {\bf 33}
 589

\bibitem{SJ96} Shannon Nic and Joyn Robert 1996 {\it J. Phys.: Condensed
Matter} {\bf 8} 10493

\bibitem{LRA73} Lee P A, Rice T M and Anderson P W 1973 {\it Phys.
Rev. Lett.} {\bf31} 463

\bibitem{RS73} Rice M J and Str"ssler S {\it Solid State Commun.} {\bf 13}
 1389

\bibitem{SGP98} Sing M, Grigoryan V G, Paasch G, Knupfer M, Fink J,
Berger H and Levy F 1998 {\it Phys. Rev.} {\bf B 57} 12768

\bibitem{McK95} McKenzie Ross H 1995 {\it Phys. Rev.} {\bf B 52} 16428

\bibitem{Sad} Sadovskii M V 1974 {\it Zh. Eksp. Theo. Fiz.} {\bf 66} 1720
(1994 {\it Sov. Phys. JETP} {\bf 39} 845); 1974 {\it Fiz. Tverd. Tela} {\bf 16}
2504 (1974 {\it Sov. Phys. Sol. State} {\bf 16} 1632); 1979
{\it Zh. Eksp. Theo. Fiz.} {\bf 77} 2070 (1979 {\it Sov. Phys. JETP} {\bf 50}
989)

\bibitem{WL76}	Wonneberger W and Lautenschl"ger R 1976 {\it J. Phys. C:
Solid State Phys.} {\bf 9} 2865

\bibitem{AL68} Aslamazov L G and Larkin A I  1968 {\it Fiz. Tverd. Tela}
{\bf 10} 1104 (1968 {\it Sov. Phys. Sol. State} {\bf 10} 875);
1968 {\it Phys. Lett.} {\bf 26A} 238

\bibitem{ST74} Str\"assler S, Toombs G A 1974 {\it Phys. Lett.} {\bf 46A} 321

\bibitem{AV93} Volkov A F 1993 {\it Solid State Commun.} {\bf 88} 715

\bibitem{AZ84} Zaitsev A V  1984 {\it  Fiz. Tverd. Tela} {\bf 26} 2669
(1984 {\it Sov. Phys. Sol. State} {\bf 26} 1618)

\bibitem{KM68} Maki K 1968 {\it Prog. Theor. Phys.} {\bf 39} 897

\bibitem{RT70} Thompson R S 1970 {\it Phys. Rev.} {\bf 1} 327

\bibitem{KM90} Maki K 1990 {\it Phys. Rev.} {\bf B 41} 9308

\bibitem{ABB74} Allender D, Bray D W and Bardeen J 1974 {\it Phys.
Rev.} {\bf B9} 119

\bibitem{FRV74} Fukuyama H, Rice T M and Varma C 1974 {\it Phys. Rev.
Lett.} {\bf 33} 305

\bibitem{PS74} Patton B R and Sham L J 1974 {\it Phys. Rev. Lett.} {\bf 33}
638; 1973 {\it Phys. Rev. Lett.} {\bf 31} 631

\bibitem{TS78} Takada S and Sakai E 1978 {\it Progr. Theor. Phys.} {\bf 59}
 1802

\bibitem{Sch78} Schulz H 1978 {\it Solid State Commun.} {\bf 34} 455

\bibitem{Ti96} Tinkham M 1996 {\it Introduction to Superconductivity}
(New York: McGraw Hill) p 308

\bibitem{RF69} Ferrel R A 1969 {\it J. Low. Temp. Phys.} {\bf 1} 241

\bibitem{JS82} Jerome D and Schulz H J 1982 {\it Adv. Phys.} {\bf 31} 299

\bibitem{DGK94} Degiorgi L, Grner G, Kim K, McKenzie R H and Wachter
P  1994 {\it Phys. Rev.} {\bf B 49} 14754

\bibitem{LRA74} Lee P A, Rice T M and Anderson P W 1974 {\it Solid State
Commun.} {\bf 14} 703

\bibitem{SLH75} Steigmeier E F, Loudon R, Harbecke G, Auderset H and
Schreiber G 1975 {\it Solid State Commun.} {\bf 17} 1447, Steigmeier E F,
Baeriswyl D, Harbecke G, Auderset H and Schreiber G 1976 ibid {\bf 20} 661

\bibitem{TMW83} Travaglini G, M\"orke I and Wachter P 1983 {\it
Solid State Commun.} {\bf 45} 289

\bibitem{CSW76} Carneiro K, Shirane G, Werner S  and Kaiser S 1976
{\it Phys. Rev.} {\bf B 13} 4258

\bibitem{PHS91} Pouget J P, Hennion B, Escribe--Filippini C and Sato
M 1991 {\it Phys. Rev.} {\bf B 43} 8421

\bibitem{FL78} Fukuyama H and Lee P A 1978 {\it Phys. Rev.} {\bf B 17} 535

\bibitem{LR79} Lee P A and Rice T M 1979 {\it Phys. Rev.} {\bf B 19} 3970

\bibitem{L87} Littlewood P B 1987 {\it Phys. Rev.} {\bf B 36} 3108

\bibitem{WW96} Wonneberger W 1996 {\it Physics and Chemistry of
Low--Dimensional Inorganic Conductors (Nato Advanced Study Insitute
at Les Houches)} ed Schlenker C et al (New York and London: Plenum) p 389

\end{thebibliography}
\end{document}